\documentclass[prl,twocolumn]{revtex4-1}

\usepackage{amsmath}
\usepackage{amssymb}
\usepackage{color}
\usepackage{amsmath} 
\usepackage{hyperref}

\begin{document}

\title{Viewpoint: Metals get an awkward cousin} 
\author{Ribhu K. Kaul}
\affiliation{Department of Physics and Astronomy, University of Kentucky, Lexington, Kentucky 40506, USA}
\begin{abstract}
A newly predicted state of matter is a simple theoretical example of a phase that conducts electricity but is not smoothly connected to our conventional model of metals.  A viewpoint on \url{http://arxiv.org/abs/1201.5998}.
\end{abstract}
\maketitle

The sustained excitement that fundamental discoveries in solid state physics have enjoyed for over a century originates at least in part from the amazing technological applications that society has found for crystalline materials that conduct electricity. This list of materials, collectively described as ``metals'' began with the familiar crystals of atoms such as Ag~\cite{ashcroft}, but has come to include complex materials like La$_{2-x}$Sr$_x$CuO$_4$, which conduct electricity like their plainer cousins, but in a manner that has vexed condensed matter physicists for decades~\cite{scho}.

Remarkably, very early on, quantum mechanics provided a simple way to understand how electrons moving in crystals can, in some cases, be insulating -- like they are in diamond; and in other cases, be metallic -- like they are in gold. Neglecting electron-electron interactions completely, the one-particle  Schr\"odinger equation in the periodic potential of a crystal has bands of allowed energies separated by bands of forbidden energies.  There are two classes of electronic structures, band-insulators in which all the bands are either fully filled or fully occupied, and band-metals in which at least one band is partially filled. Fully occupied or fully empty bands are electrically inert and it is the partially filled bands that endow metals with the characteristic property of electrical conduction. The non-interacting picture of a band-metal can be improved upon by including the interactions of electrons with each other assuming they are weak. This dressed-up version of the original non-interacting picture of metallicity is called the ``Fermi liquid'' phenomenology. Despite the ad hoc assumption that the electron-electron interactions can be treated as a small perturbation, the ``Fermi-liquid'' picture makes a number of predictions that have been borne out in experiment. It has hence enjoyed the status of being the exclusive paradigm used to interpret the low temperature properties of metals in the 20$^{th}$ century~\cite{abrikosov}.

But what if electron-electron interactions are so strong as to make the band theory an incorrect starting point? Towards the end of the first half of the 20$^{\rm th}$ century, Neville Mott and others proposed that electrons in {\em all} crystals, driven by their interactions with each other, would eventually become insulating if the lattice spacing could artificially be made large enough~\cite{mott_book}. It turns out that Mott's picture was not a theorists' pipe dream: there are hundreds of materials in such a regime that electron motion is inhibited completely by electron-electron interactions at low temperatures resulting in insulating behavior. Such materials, called Mott insulators, can have exotic magnetic properties despite their inability to conduct electricity. The study of quantum and classical magnetism in Mott insulators has evolved into one of the hottest frontiers of both experimental and theoretical condensed matter physics in recent decades~\cite{leon_nature}.
 
Given the exciting many-body physics in Mott insulators, it is natural to ask whether new ``non-Fermi liquid'' paradigms for metallic states can emerge from the complexity of electron-electron interactions. In this field, experiments are far ahead of theories. More than a decade ago, Greg Stewart compiled in excess of 50 examples of metals that have properties that deviate from the ``Fermi liquid'' phenomenology~\cite{stewart}. It is generally accepted that this new behavior is a result of electron-electron interaction and perhaps disorder. Despite the high stakes, there is not yet a universally accepted simple paradigm that has enabled an understanding of these deviant behaviors~\cite{qc}. While an explanation of experimental observations is the ultimate goal of theories, as a first step, simple theoretical models which have metallic ground states and yet do not fall into the Fermi-liquid/band theory paradigm are highly desirable. 
 
In the current issue of Physical Review B, Nandkishore, Metlitski and Senthil describe a ``non-Fermi liquid'' phase of matter that they have christened the ``orthogonal metal.'' The name originates from the interesting property that the wavefunctions of the current carrying fermions in this phase are orthogonal to the microscopic electrons, in contrast to the usual Fermi liquids where the two have a strong overlap. The orthogonality feature leads to a state of matter that conducts electricity much like an ordinary metal but even so has distinct and unusual spectral properties that earn it the title of a ``non-Fermi liquid.''  The orthogonal metal is especially worth taking note of, since it is one of the simplest theoretical examples of a non-Fermi liquid metal.

Technically, one procedure to access the ``orthogonal metal'' phase is by writing the microscopic lattice electron operator, $c_{i\sigma}$, on the site $i$ of a lattice as a product of a fermion operator, $f_{i\sigma}$, and an Ising spin, $\tau^x_i$,
\begin{equation}
\label{eq:slave}
c_{i\sigma} = f_{i\sigma}\tau^x_i.
\end{equation}
This re-writing of the $c$-electron supplemented by a local constraint on the $f$ and $\tau$ degrees of freedom is referred to as the slave spin construction. It is an efficient way to access unconventional phases of the electrons beginning with well understood phases of the $f$ and $\tau$ particles. A central clue to the interpretation put forward in this work is that in the break-up of the electron of Eq.~(\ref{eq:slave}), the $f$ fermions carry both the electrical charge and spin of the electron. This finding has significant consequences for the phases that can be accessed simply in the new representation, since it implies that in any phase in which the $f$-fermions form a metal, so will the microscopic $c$-electrons. Indeed the authors show that the phase in which the $\tau$ particle is condensed corresponds to a conventional Fermi liquid in which there is a finite overlap between the $f$ and the microscopic $c$-electrons. They then go on to show that when the $\tau$ particles are disordered one obtains the ``orthogonal metal'' in which the $f$ fermions have no overlap with the original microscopic $c$-electrons. The most striking feature of this new phase of matter is that it is compressible like an ordinary metal, yet its spectral function has a gap because of the fractionalization of the $c$-electron.

The slave particle method is a convenient way to envision {\em possible} exotic phases of $c$-electrons. However, a weakness of the slave particle approach is that it is unable to identify specific microscopic models of the $c$-electrons that realize the new phase. The authors address this important issue by providing exactly soluble ``toy'' quantum Hamiltonians in which the new phase is realized as a ground state.  These exactly soluble models are not the typical ``generic'' Hamiltonians physicists would use to model condensed matter systems, yet they provide important ``proof of principle'' examples that this new exotic phase can be stabilized in models with simple local interactions.

Looking forward, it is too early to judge how this new phase of matter will assist in the interpretation of the mysterious experimental data on the deviant metallic materials mentioned before. Its existence should certainly be kept in mind in looking for possible explanations of metals that display non-Fermi liquid behavior. Detection may not always be obvious since outside of a few measurements that probe one electron properties, such as photo-emission and tunneling, ``orthogonal metals'' look very much like ordinary metals in most experiments.  An intriguing, although admittedly optimistic possibility is that we have already synthesized materials which are ``orthogonal metals,'' but have mistaken them for ordinary metals because of the subtle differences between these two phases!
An immediate challenge for theoretical work is to find  ``generic'' microscopic models of electrons that realize the ``orthogonal metal'' in their phase diagrams. To this end, the authors have provided variational wavefunctions as well as likely ``generic'' models that could realize this new phase.   Recent advances in numerical methods for lattice Hamiltonians have allowed unbiased studies of many new models that harbor exotic critical points and phases~\cite{kms}, making one hopeful that the existence of orthogonal metals can be convincingly demonstrated in a generic model reasonably soon in the future. It is clearly an exciting direction for future computational and experimental work to find for the new  ``orthogonal metal,''  a path from exotica to the catalogue of standard condensed matter phases.

The work was supported in part by NSF DMR-1056536.

\end{document}